# A numerical solution to the minimum-time control problem for linear discrete-time systems

Laurent Bako, Dulin Chen and Stéphane Lecoeuche

(1) - Univ Lille Nord de France, F-59000 Lille, France
(2) - EMDouai, IA, F-59500 Douai, France
{laurent.bako,dulin.chen,stephane.lecoeuche}@mines-douai.fr

**Abstract**

The minimum-time control problem consists in finding a control policy that will drive a given dynamic system from a given initial state to a given target state (or a set of states) as quickly as possible. This is a well-known challenging problem in optimal control theory for which closed-form solutions exist only for a few systems of small dimensions. This paper presents a very generic solution to the minimum-time problem for arbitrary discrete-time linear systems. It is a numerical solution based on sparse optimization, that is the minimization of the number of nonzero elements in the state sequence over a fixed control horizon. We consider both single input and multiple inputs systems. An important observation is that, contrary to the continuous-time case, the minimum-time control for discrete-time systems is not necessarily entirely bang-bang.

## 1 Introduction

Time optimal control is an important particular instance of the general theory of optimal control developed since the 50's. It refers to the problem of transferring the state of a dynamic system from a given initial state to a certain target state (or a set of states) in minimum time. This problem can be encountered in many military applications such as the interception of an attacking missile or in mobile robotics applications. Attempting to provide a closed-form solution to the general minimum-time control problem is a daunting task that can get quickly hopeless. For this reason, the only existing analytic solutions concern simple continuous-time examples of small dimensions [1, 15, 6]. Even in these simple cases, the problem proves to be extremely hard to tackle. For continuous-time systems, the minimum-time problem is generally addressed by applying the minimum/maximum principle of Pontryagin [1]. When the control variables are bounded, a general result is that the optimal control policy is a bang-bang policy that is a policy in which the control switches between a number of extreme values. In discrete-time systems, results on minimum-time problem are rather very scarce. Only a few papers [11, 17, 3, 14] investigated the problem. In a more recent work reported in [12], a closed-form optimal policy is derived for the well-known double integrator system. Another relevant work is the one of [20], where an intricate derivation of a minimum-time control for a chain of three integrators is presented. Inspired by the book [18], the author of [12] showed on the particular double integrator system that contrarily to the continuous-time case studied in [1, 6], the solution to the minimum-time problem with bounded controls, is not necessarily entirely bang-bang for discrete-time systems. The results of this paper will, even though they are not intended for characterizing the form of the optimal-time control law, confirm this intriguing fact.

**Contributions.** In this paper, we consider the problem of steering the state of a linear discrete-time system from a given nonzero state to the zero-state in minimum time (and maintain it at zero at all subsequent instants). The controls to be selected for achieving this goal are constrained to lie in a bounded convex set. The paper develops a new, generic and computationally feasible approach to the minimum-time problem for discrete-time linear systems. Our approach is based on a sparse optimization of the state sequence, that is the minimization of the number of nonzero elements contained in the state sequence over a fixed and finite control horizon. In effect, the state sequence associated with the time-optimal control policy can be viewed as the sparsest state trajectory achievable by the considered system.[1]

Sparse optimization refers to the general problem of minimizing the number of nonzero elements in a sequence of scalars or a sequence of vectors. It has proved to be a powerful method having impressive applications in many different engineering areas such as compressed sensing [8], signal processing [5], computer vision [19], system identification [2]. Inspired by these results, we show in this paper that sparsity-inducing optimization techniques can efficiently solve the minimum-time control problem.

Note however that a direct attempt to solve the sparse optimization problem is computationally very intensive and somewhat inelegant. We therefore approximate it by minimizing a cost functional constructed as an appropriately weighted sum of $\ell_2$-norms of the system states. This relaxation of the minimum-time problem has the advantage of being convex and therefore solvable by some reliable numerical tools [4, 13]. Sufficient conditions are derived for the relaxed convex optimization to exactly recover the solution to the minimum-time control problem. In contrast to many existing methods which studied particular examples, our method applies to arbitrary linear systems.

**Outline.** The rest of the paper is organized as follows. The minimum-time control problem is formally set up in Section 2 along with some working assumptions. Some preliminary analysis of the solution is also carried out. The proposed approach is developed in Section 3. Section 4 contains a numerical illustration of the capability of our approach. Final remarks are provided in Section 5.

---

[1] A vector sequence is said to be sparse if it contains many zero vectors. And a vector sequence is sparser than another vector sequence if the first contains a larger number of zeros than the second. As for the paradigm sparse optimization, it is used here to designate an optimization problem which aims at minimizing an integer such as the cardinality of a set, the rank of matrix, ...

## 2  The minimum-time control problem

Consider the discrete-time linear time-invariant (LTI) system represented by

$$x(t+1) = Ax(t) + Bu(t), \quad x(0) = x_0 \tag{1}$$

where $t \in \mathbb{Z}_{\geq 0}$ refers to time step, $x(t) \in \mathbb{R}^n$ is the state, $u(t) \in \mathbb{R}^{n_u}$ is the input. $A \in \mathbb{R}^{n \times n}$ and $B \in \mathbb{R}^{n \times n_u}$ are the system matrices. The objective of this work is to provide a solution to the following control problem.

**Problem 1.** *Given a linear discrete-time system such as (1) with the initial nonzero state $x(0) = x_0$, find the control sequence $u(0), \ldots, u(T^* - 1)$, subject to the constraints*

$$u(t) \in \mathcal{U}, \quad t = 0, \ldots, T^* - 1$$

*such that the system state is driven to the zero-state $x(T^*) = 0$ in minimum time. That is, the integer $T^*$ must be the minimum time at which the system reaches the zero-state. $\mathcal{U} \subset \mathbb{R}^{n_u}$ represents a bounded convex set and will be referred to as the set of admissible controls. It is additionally required that $\mathcal{U}$ contains the origin (the zero input).*

More formally, the control problem to be studied in this paper reads as follows

$$\begin{aligned}
&\min_{u(0),\ldots,u(T-1)} T \\
&\text{subject to Eq. (1)} \\
&x(T) = 0 \text{ and } x(t) \neq 0, \ t = 0, \ldots, T-1, \\
&u(t) \in \mathcal{U}, \ t = 0, \ldots, T-1.
\end{aligned} \tag{2}$$

In the setting of this paper, $\mathcal{U}$ can be any convex set in $\mathbb{R}^{n_u}$. In particular, $\mathcal{U}$ can be an $\ell_2$-norm ball of the form $\{u \in \mathbb{R}^{n_u} : \|u\|_2 \leq r\}$ or an $\ell_\infty$-norm ball $\{u \in \mathbb{R}^{n_u} : |u_i| \leq r_i, \ i = 1, \ldots, n_u\}$. We assume here that the initial state $x_0$ is nonzero and the target state is equal to zero. An implicit requirement is that the system must stay in the zero-state after it reaches it. Obviously, for the above minimum-time control problem to make sense, we need to assume that the zero-state is reachable in finite time from $x_0$, under the control constraints $u(t) \in \mathcal{U}$. The problem would be trivial if the control variables were not constrained i.e., if $\mathcal{U}$ were equal to $\mathbb{R}^{n_u}$ because the minimum-time $T^*$ could, in this case, be simply obtained as $T^* = \min\{t : A^t x_0 \in \text{im}(\Delta_t)\} \leq n$

with
$$\Delta_t = \begin{bmatrix} A^{t-1}B & \cdots & AB & B \end{bmatrix}, \qquad (3)$$

$\text{im}(\Delta_t)$ referring to the range space of $\Delta_t$. Many solutions would then be possible once $\Delta_{T^*}$ is a rectangular matrix that is, once $n_u T^* > n$. The case of real interest is the one where $\mathcal{U}$ is sharply bounded. In this case the minimum-time is formally given by

$$T^* = \min \left\{ t : \mathcal{U}^t \cap F_{\Delta_t}(-A^t x_0) \neq \emptyset \right\} \qquad (4)$$

where $\mathcal{U}^t = \mathcal{U} \times \cdots \times \mathcal{U}$ is the $t$-cartesian product of the set $\mathcal{U}$ and $F_{\Delta_t}(z) = \left\{ \bar{u} \in \mathbb{R}^{tn_u} : \Delta_t \bar{u} = z \right\}$. Note that $T^*$ depends on the system (through its matrices), the control set $\mathcal{U}$ and the initial state $x_0$. Also, it is worth noting from (4) that $\mathcal{U}^{T^*} \cap F_{\Delta_t}(-A^{T^*} x_0)$ is the set of time-optimal control sequences. Since this set is convex, we see that whenever the time-optimal control sequence is not unique, there are infinitely many control sequences that can achieve the minimum time.

For single input continuous-time systems, it is well-known from the optimal control literature that the solution to the minimum-time problem is, when the controls are bounded, a bang-bang control policy. A bang-bang control policy is the one in which the control switches between two extreme values, for example $u(t) = \pm 1$ for $\mathcal{U} = \{u \in \mathbb{R} : |u| \leq 1\}$. For discrete-time systems however the control policy is not necessarily bang-bang [12].

Note that posed as in (2), the minimum-time problem is a hard mixed integer-continuous variables optimization problem. In the sequel, we propose a convex relaxation which is more numerically tractable.

## 3 The sparse optimization based solution

### 3.1 Description of the approach

In this section, we present the main contribution of the paper. It consists in the development of a new convex optimization based method for computing the solution to the minimum-time control problem. To proceed with the presentation, denote by $T^*$ the minimum possible time in which the zero target state can be reached from initial state $x_0$ and let $T$ be a known upper bound of $T^*$, i.e. $T^* \leq T$. This assumption will remain in force throughout the paper. Let $x^*(\cdot)$ be a (many of them may exist) time-optimal state trajectory. Then by fixing the control horizon to be equal to $T$, the state sequence $x^*(0), x^*(1), \ldots, x^*(T^* - 1), x^*(T^*), \ldots, x^*(T)$ is such that $x^*(t) \neq 0$ for $t < T^*$ and $x^*(t) = 0$ for $t \geq T^*$. The fundamental idea of our approach is based on the important observation that for $x^*(\cdot)$ to be the state trajectory of system (1) corresponding to the minimum-time control policy, the vector sequence $\{x^*(t)\}$ must be one of the sparsest achievable state trajectories. That is, the sequence $\{x^*(t)\}$ must contain as many zero vectors as possible with all zero states corresponding to the highest values of the time index $t$. This feature of the state sequence $\{x^*(t)\}$ is graphically depicted as follows

$$
\begin{array}{cccccccc}
x^*(1) & x^*(2) & \cdots & x^*(T^*-1) & x(T^*) & \cdots & x(T) \\
\downarrow & \downarrow & \cdots & \downarrow & \downarrow & \cdots & \downarrow \\
\# & \# & \cdots & \# & 0 & \cdots & 0
\end{array}
$$

where $\#$ means nonzero. In fact, requiring the sequence $\{x^*(t)\}$ of state vectors to be sparse is, in principle, equivalent to making sparse the scalar sequence $\{\|x^*(t)\|_2\}$, where $\|\cdot\|_2$ refers to the euclidean norm. We therefore reformulate problem (2) as the problem of minimizing over the horizon $T$, the number of nonzero elements in the sequence $\{\|x^*(t)\|_2\}$,

$$
\begin{aligned}
&\min_{u(0),\ldots,u(T-1)} |\{t : \|x(t)\|_2 \neq 0\}| \\
&\text{subject to } \text{Eq. (1)} \\
&\quad \exists\, T_1 \leq T : x(t) = 0,\ t = T_1,\ldots,T \\
&\quad \text{and } x(t) \neq 0,\ t = 0,\ldots,T_1 - 1, \\
&\quad u(t) \in \mathcal{U},\ t = 0,\ldots,T-1.
\end{aligned}
\qquad (5)
$$

Here, the notation $|\mathcal{S}|$ with $\mathcal{S}$ representing a set, stands for the cardinality of the set $\mathcal{S}$. Like (2), problem (5) is a hard combinatorial problem. Both problems can be obviously solved in a somewhat inelegant way by trying any possible value for $T_1$ (starting from $T_1 = 1$) until the minimum time is obtained, i.e. until the constraints associated with problem (5) are satisfied. This comes however with a huge numerical complexity as it involves solving an a priori undefined number of times a constrained control problem over a time horizon $T$, which $T$ is possibly large.

A more effective way to tackle problem (5) is to replace it by its "best" convex approximation. This translates into the formulation

$$
\begin{aligned}
&\min_{u(0),\ldots,u(T-1)} J\big(\{u(k)\}_{k=0}^{T-1}\big) = \sum_{t=1}^{T} w(t)\,\|x(t)\|_2 \\
&\text{subject to } \text{Eq. (1)} \\
&\quad u(t) \in \mathcal{U},\ t = 0,\ldots,T-1
\end{aligned}
\qquad (6)
$$

where $\{w(t)\}$ is a strictly increasing sequence of positive weights. Noting that

$$
x(t) = A^t x(0) + \Delta_t \bar{u}_t \qquad (7)
$$

with $\Delta_t$ defined as in (3) and

$$
\bar{u}_t = \begin{bmatrix} u(0)^\top & \cdots & u(t-1)^\top \end{bmatrix}^\top, \qquad (8)
$$

the cost functional $J$ in (6) can be expressed in function of the control variables as

$$J\big(\{u(k)\}_{k=0}^{T-1}\big) = \sum_{t=1}^{T} w(t) \left\| A^t x(0) + \Delta_t \bar{u}_t \right\|_2$$

with $\Delta_t$ and $\bar{u}_t$ defined respectively as in (3) and (8). The motivation for the optimization problem (6) is that the sum-of-$\ell_2$-norms criterion has the property of promoting the obtention of a control sequence such that, as desired, the state sequence $\{x(t)\}$ is as sparse as possible, see e.g. [10, 16] for some discussions. The non-smooth sum-of-$\ell_2$-norms criterion is in fact a generalization of the $\ell_1$-norm which is popular in the literature of compressed sensing [9, 5] for its ability to convexly approximate the combinatorial $\ell_0$-norm. A clear distinction needs to be made between the sum-of-norms cost functional involved in (6) and the more usual sum-of-squared-norms. The former is non-smooth, non-differentiable at zero and tends to stimulate the obtention of a sparse state sequence. The latter is smooth and differentiable, so that it does not favor sparse solutions. The increasing sequence $\{w(t)\}$ of weights aims at forcing faster (non-smoothly) the states $x(t)$ to zero as $t$ is increasing.

Note that the formulation (6) is a convex approximation of the initial minimum-time control problem (2). Problem (6) can, contrary to (2) and (5), be efficiently solved using some reliable numerical solvers such as e.g., the CVX toolbox [13, 4]. However we may wonder under what conditions the solutions of both optimization problems coincide. In other words, can the time-optimal problem be exactly tackled through convex optimization?

An answer is provided by Theorems 1 and 2. For ease of presentation of the first theorem, we need to introduce a few more notations. For any finite input sequence $u(\cdot) = \{u(0), \ldots, u(T-1)\}$, let $\bar{u}_t$, $t \leq T$, be defined as $\left[u(0)^\top, \ldots, u(t-1)^\top\right]^\top \in \mathbb{R}^{tn_u}$. We call $\mathbb{U}_T(x_0)$ the set of all input sequences of length $T$ which solve problem (6) when the system starts from state $x_0$ that is,

$$\mathbb{U}_T(x_0) = \Big\{ u^o(\cdot) = \{u^o(0), \ldots, u^o(T-1)\} \subset \mathcal{U} :$$
$$J(u^o(\cdot)) \leq J(u(\cdot)) \ \forall \ u(\cdot) \subset \mathcal{U} \Big\}.$$

$\mathbb{U}_T(x_0)$ is in fact a convex set so that it is, unless reduced to a singleton, an infinite uncountable set. Now denote by $\mathbb{U}_T^*(x_0)$ the set of all truncated sequences (elements) of the form $\bar{u}_t$ taken from $\mathbb{U}_T(x_0)$ and define

$$\eta_T(x_0) = \min_{\substack{\bar{u}_t \in \mathbb{U}_T^*(x_0) \\ A^t x_0 + \Delta_t \bar{u}_t \neq 0}} \left\| A^t x_0 + \Delta_t \bar{u}_t \right\|_2. \tag{9}$$

The so-defined $\eta_T(x_0)$ is a lower bound for any non-zero state $x(t)$ which is reachable by a control sequence $\bar{u}_t \in \mathbb{U}_T^*(x_0)$. Hence $\|x(t)\|_2 \geq \eta_T(x_0)$ whenever the state $x(t)$ generated by $\bar{u}_t \in \mathbb{U}_T^*(x_0)$ is different from zero.

**Theorem 1.** *Fix the control horizon to satisfy $T^* \leq T$ and assume that the set $\mathcal{U}$ of admissible controls*

is bounded by a number $r > 0$. If the weights sequence $\{w(t)\}$ are recursively constructed as

$$\begin{cases} w(1) = 1, \\ w(t) > \dfrac{2r}{\eta_T(x_0)} \sum_{k=1}^{t-1} \sqrt{k}\, \|\Delta_k\|_2\, w(k) \text{ for } t \geq 2, \end{cases} \quad (10)$$

then the convex problem (6) provides a solution to the minimum-time control problem (2).

Before proving the theorem, we make the following observation.

**Lemma 1.** *Assume that the weight sequence $\{w(t)\}$ is strictly positive. Then by denoting with $u^o(\cdot)$ and $x^o(\cdot)$ respectively the control and state trajectories corresponding to a solution to problem (6), the following holds. If $x^o(t_0) = 0$ for some $t_0 \in \{1, \ldots, T\}$, then $x^o(t) = 0$ for any $t$ such that $t_0 \leq t \leq T$.*

*Proof.* Define a control sequence $u^1(\cdot)$ as $u^1(t) = u^o(t)$ for $t \leq t_0 - 1$ and $u^1(t) = 0$ for $t \geq t_0$. Define $x^1(\cdot)$ to be the state sequence induced by the control sequence $u^1(\cdot)$. Then $x^1(t) = x^o(t)$ for any $t \leq t_0$ and $x^1(t) = 0$ for any $t \geq t_0$. Since $u^o(\cdot)$ minimizes the cost function in (6), we can write

$$\sum_{t=1}^{T} w(t) \|x^o(t)\|_2 \leq \sum_{t=1}^{T} w(t) \|x^1(t)\|_2$$

which implies that

$$\sum_{t=t_0+1}^{T} w(t) \|x^o(t)\|_2 \leq 0.$$

This, by virtue of the fact that $w(t) > 0$ for any $t$, is true only if $x^o(t) = 0$ for any $t = t_0 + 1, \ldots, T$. □ □

The lemma implies that if an instance of problem (6) solves the minimum-time problem with a control horizon $T \geq T^*$, then the solutions to all instances of (6) having a control horizon $T' \geq T^*$, also achieve the minimum time.

*Proof of Theorem 1.* Use the notations $u^o(\cdot)$ and $x^o(\cdot)$ to represent respectively the control and state trajectories corresponding to the solution to problem (6). We just need to show that if there is a control policy achieving the minimum-time $T^*$, then $x^o(\cdot)$ satisfies necessarily $x^o(t) = 0$ for all $t \geq T^*$. For this purpose, denote with $u^*(\cdot)$ the minimum-time control trajectory and with $x^*(\cdot)$ the associated state trajectory. From the definition of $x^o(\cdot)$, it holds that

$$\sum_{t=1}^{T} w(t) \|x^o(t)\|_2 \leq \sum_{t=1}^{T} w(t) \|x^*(t)\|_2.$$

Considering the fact that $x^*(t) = 0\ \forall t \geq T^*$ and using formula (7), the above equality implies

that

$$\sum_{t=T^*}^{T} w(t)\left\|x^o(t)\right\|_2 \leq \sum_{t=1}^{T^*-1} w(t)\left\|x^*(t)\right\|_2 - \sum_{t=1}^{T^*-1} w(t)\left\|x^o(t)\right\|_2$$

$$\leq \sum_{t=1}^{T^*-1} w(t)\left\|x^*(t) - x^o(t)\right\|_2$$

$$= \sum_{t=1}^{T^*-1} w(t)\left\|\Delta_t(\bar{u}_t^* - \bar{u}_t^o)\right\|_2.$$

Because $\bar{u}_t^o$ and $\bar{u}_t^*$ are included in the bounded set $\mathcal{U}$, the following inequality holds

$$w(t)\left\|\Delta_t(\bar{u}_t^* - \bar{u}_t^o)\right\|_2 \leq 2rw(t)\left\|\Delta_t\right\|_2 \sqrt{t}$$

so that we arrive at

$$\sum_{t=T^*}^{T} w(t)\left\|x^o(t)\right\|_2 \leq \sum_{t=1}^{T^*-1} 2rw(t)\left\|\Delta_t\right\|_2 \sqrt{t}.$$

If there were a $T_1$ in $\{T^*, \ldots, T\}$ such that $x^o(T_1) \neq 0$, then by Lemma 1, $x^o(T^*) \neq 0$. We would then have

$$\eta_T(x_0) w(T^*) \leq w(T^*)\left\|x^o(T^*)\right\|_2 \leq \sum_{t=1}^{T^*-1} 2rw(t)\left\|\Delta_t\right\|_2 \sqrt{t}.$$

However, this is incompatible with the definition (10) of the weights $\{w(t)\}$. Therefore $x^o(t) = 0$ for any $t \geq T^*$. In other words, the control $u^o(\cdot)$ is a solution to the minimum-time problem as claimed. □ □

Theorem 1 provides us with a sufficient condition on the weights sequence for a minimum-time control to be obtained by means of convex optimization. While condition (10) offers a theoretical expedient for problem (6) to solve the minimum-time problem, it may not be of great practical significance if the control horizon $T$ is large. The essential reason for that is the undesirable fact that the weights may grow too rapidly. This is particularly critical when the considered system is unstable. Having too large weights $w(t)$ for big values of $t$ may have the numerical inconvenience of annihilating the terms of the cost functional indexed by small time indexes since the corresponding weights will be too small. In fact, it has turned out in practice that the condition (10) is somewhat conservative. A weight sequence which increases linearly with respect to time is in general capable of providing an optimal solution. Another issue related to how to fulfill condition (10), is that the parameter $\eta_T(x_0)$ may be difficult to compute in practice. One can get around this difficulty by just picking $\eta_T(x_0)$ to be sufficiently small. Note that a subsequent normalization of the weights is possible.

The next theorem gives a sufficient condition under which solving problem (6) leads uniquely to the minimum-time control sequence.

**Theorem 2.** *For any control sequence $u(\cdot)$ and any integer $T \geq 1$, define*

$$\rho_T(u(\cdot)) = \sum_{t=1}^{T} w(t) \, \|\Delta_t \bar{u}_t\|_2 \,.$$

*Denote with $\mathcal{U} \ominus \mathcal{U}$ the Minkowski difference[2] of the set $\mathcal{U}$ from itself and introduce, for any $T_1 \leq T$, the number*

$$\mu_{\mathcal{U}}(T_1, T) = \sup_{\substack{u(\cdot) \subset \mathcal{U} \ominus \mathcal{U} \\ \rho_T(u(\cdot)) \neq 0}} \frac{\rho_{T_1}(u(\cdot))}{\rho_T(u(\cdot))}. \tag{11}$$

*Assuming that the weight sequence $\{w(t)\}$ has strictly positive elements, the following holds. If there is $T_1$, $T^* - 1 \leq T_1 \leq T$, such that*

$$\mu_{\mathcal{U}}(T_1, T) < 1/2, \tag{12}$$

*then the solution to problem (6) achieves the minimum time control. If in addition $\mathrm{rank}(B) = n_u$, the control sequence achieving the minimum time is unique.*

*Proof.* Let us adopt the same notations as in the proof of Theorem 1. From that proof, we know that

$$\sum_{t=T^*}^{T} w(t) \, \|x^o(t)\|_2 \leq \sum_{t=1}^{T^*-1} w(t) \, \|\Delta_t(\bar{u}_t^o - \bar{u}_t^*)\|_2 \,.$$

Because $u^*(t) = 0$ for any $t \geq T^*$, we have $x^*(t) = A^t x_0 + \Delta_t \bar{u}_t^* = 0$ so that $A^t x_0 = -\Delta_t \bar{u}_t^*$ for $t \geq T^*$. As a consequence, $x^o(t)$ can be expressed as $x^o(t) = \Delta_t(\bar{u}_t^o - \bar{u}_t^*) \; \forall \, t \geq T^*$. Plugging this into the above inequality yields

$$\sum_{t=1}^{T} w(t) \, \|\Delta_t \bar{u}_t\|_2 \leq 2 \sum_{t=1}^{T^*-1} w(t) \, \|\Delta_t \bar{u}_t\|_2 \tag{13}$$

where $\bar{u}_t = \bar{u}_t^o - \bar{u}_t^*$, $u(\cdot)$ being the sequence defined by $u(\cdot) = u^o(\cdot) - u^*(\cdot) \subset \mathcal{U} \ominus \mathcal{U}$. If $\rho_T(u(\cdot)) \neq 0$, then inequality (13) implies that

$$\mu_{\mathcal{U}}(T_1, T) \geq \frac{\rho_{T^*-1}(u(\cdot))}{\rho_T(u(\cdot))} \geq 1/2$$

However, this is incompatible with the assumption (12) of the theorem. Therefore it necessarily holds that $\rho_T(u(\cdot)) = 0$. With $w(t) > 0$ for any $t$, this immediately leads to the conclusion that $\Delta_t(\bar{u}_t^o - \bar{u}_t^*) = 0$ for any $t = 1, \ldots, T$. In particular,

$$x^o(T^*) = A^{T^*} x_0 + \Delta_{T^*} \bar{u}_{T^*}^o = \Delta_{T^*}(\bar{u}_{T^*}^o - \bar{u}_{T^*}^*) = 0.$$

---

[2]The Minkowski difference of two sets $A$ and $B$ is the set $A \ominus B$ defined by $A \ominus B = \{a - b : a \in A, b \in B\}$.

In other words, the control sequence $u^o(\cdot)$ achieves the minimum time.

Uniqueness of the solution is immediate by writing explicitly the equations $\Delta_t \bar{u}_t = 0$ for $t = 1, \ldots, T$. We get

$$Bu(0) = 0,$$
$$ABu(0) + Bu(1) = 0,$$
$$A^2 Bu(0) + ABu(1) + Bu(2) = 0,$$
$$\cdots \quad \cdots \quad \cdots.$$

From these equations, it is clear when $\mathrm{rank}(B) = n_u$, that $u(\cdot) = u^o(\cdot) - u^*(\cdot) = 0$ and so, $u^o(\cdot) = u^*(\cdot)$. Since $u^*(\cdot)$ represents any sequence that realizes the minimum time, we can conclude on the uniqueness of the minimum-time control sequence. □ □

It is important to notice that (12) is a property of the system, the set $\mathcal{U}$ of admissible controls and the initial state $x_0$. The dependence on the initial state is hidden in the minimum $T^*$. Although condition (12) may be a little difficult to check in practice, it does consolidate the idea that, by choosing the weights sequence to be appropriately increasing, the minimum-time control sequence can be computed via convex optimization. And, the reasonably larger the control horizon $T$ is, the more likely condition (12) is to hold.

**Remark 1.** *The method of this paper has been presented only for the case where the target set is reduced to a zero singleton. In fact, it can be directly extended to cover the situations where the target is an invariant set of the linear transformation $x \mapsto Ax$. Such a set is a subspace which equals the null space of $A - I_n$ with $I_n$ being the identity matrix of order $n$.*

**Remark 2.** *Note that we can, as in [14], impose on both the state and the control input, some linear constraints of the form $(x(t), u(t)) \in \mathcal{Z}$ where*

$$\mathcal{Z} = \{(x, u) \in \mathbb{R}^n \times \mathbb{R}^{n_u} : Ex + Fu + \lambda \leq 0\}$$

*where $E \in \mathbb{R}^{l \times n}$, $F \in \mathbb{R}^{l \times n_u}$, $\lambda \in \mathbb{R}^l$ with $l$ being the number of inequalities. The inequality symbol in the definition of $\mathcal{Z}$ should be understood component-wise. Imposing such constraints leaves the problem (6) still convex.*

## 3.2 Online implementation

The control approach presented above can be implemented online in a similar fashion as the so-called Model Predictive Control (MPC) [7]. At time $t$, $x(t)$ is known and $u(t)$ is obtained as $u(t) = u(t|t)$, where $u(t|t)$ is the first value of the solution $u(t|t), \ldots, u(t + \tau - 1|t)$ of the

optimization problem

$$\min_{u(t|t),\ldots,u(t+\tau-1|t)} \sum_{k=1}^{\tau} w(t+k) \, \|x(t+k|t)\|_2$$
$$\text{subject to Eq. (1)}$$
$$u(t+k|t) \in \mathcal{U}, \ k=0,\ldots,\tau-1. \tag{14}$$

Here, $\tau$ is a time horizon that need not be larger than the minimum time $T^*$. It should however be chosen greater than the order $n$ of the system. The control $u(t|t)$ computed from (14) can be viewed as a function of the state $x(t)$. Hence by denoting $u(t|t) = \mu(x(t))$, the online minimum-time control policy can be written as

$$u(t) = \begin{cases} \mu(x(t)) & \text{if } x(t) \neq 0 \\ 0 & \text{otherwise}. \end{cases}$$

Note that it is not necessary to solve problem (14) at each sampling time. This can in fact be done periodically. For instance one can apply, without additional calculations, the controls $u(t|t), \ldots, u(t+\tau-1|t)$ computed at time $t$ to the system, during the period $[t, \ t+\tau-1]$. Hence problem (14) can be solved periodically at each time $q\tau$, $q \in \mathbb{Z}_{\geq 0}$.

## 4 Numerical results

### 4.1 A double integrator system

A system that has been extensively used in the literature of time-optimal control of continuous-time systems is the double integrator system. We found it appropriate to also test our approach on that system as this will allow for comparison with the closed-form solution derived in e.g., [12]. The double integrator is described by a model of the form (1) with

$$A = \begin{bmatrix} 1 & T_e \\ 0 & 1 \end{bmatrix}, \quad \text{and} \quad B = \begin{bmatrix} 0 \\ T_e \end{bmatrix} \tag{15}$$

where $T_e$ is the sampling time which is assumed to be equal to $1$. Starting from a nonzero initial state, we are interested in determining a control sequence that can drive the double integrator to the zero state in the minimum possible time. The control constraint set is defined by $\mathcal{U} = \{u \in \mathbb{R} : |u| \leq 1\}$. The control horizon is set to $T = 10$ and the weights are selected to be linearly increasing i.e., $w_i(t) = at$, with $a > 0$. The proposed algorithm is then run when the system starts from a certain number of different initial states. The results for the optimal state trajectories are plotted in Figure 1. An immediate salient observation here is that the obtained trajectories are, like in the continuous-time case, switching among parabolic curves whose axis of symmetry is the x-axis. However, because the sampling time is relatively large, the parabola seem to have been made of segments of lines. The smaller $T_e$ is, the smoother the trajectories become. Not surprisingly, setting $T_e$ small will require more steps to reach the desired zero-state. One fundamental difference with the continuous-time control is, as already mentioned, that the solution to problem (2) is not necessarily a bang-bang control (see Figure 2).

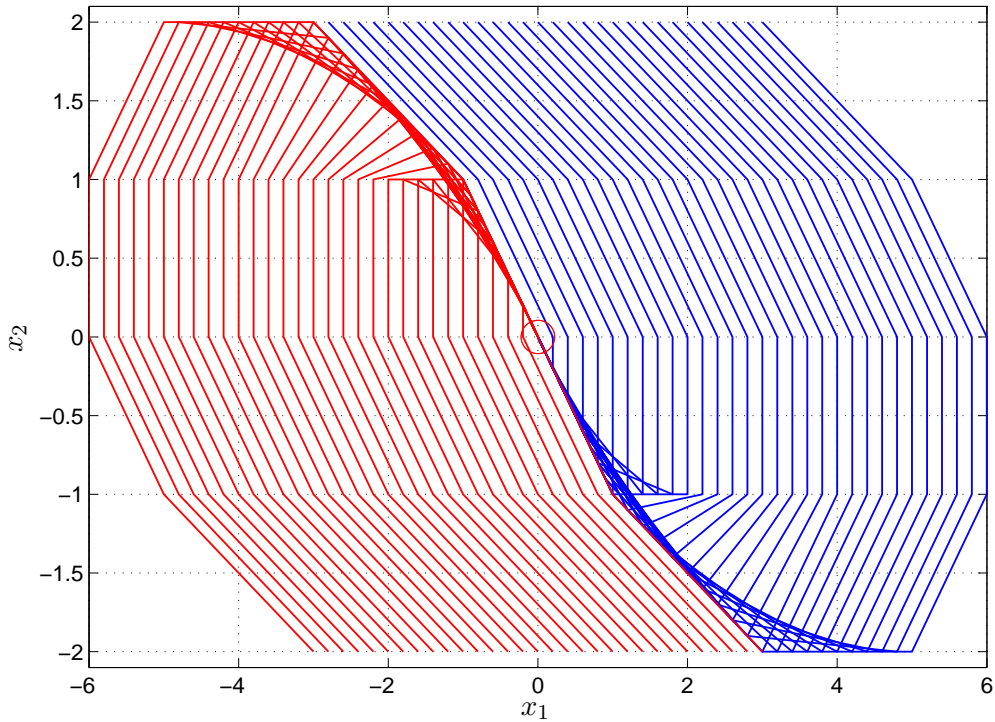

Figure 1: Optimal trajectories of the state corresponding to the minimum-time problem.

### 4.2 A multi-input system

To further illustrate the applicability of our method, we now consider a multivariable linear system generated at random with 2 inputs and state dimension equal to 3. The corresponding state-space model is described in the form (1) by

$$A = \begin{bmatrix} -0.093 & 0.25 & 0.500 \\ -0.540 & -0.255 & 0.160 \\ -0.072 & 0.525 & -0.445 \end{bmatrix}, \text{ and } B = \begin{bmatrix} 0.580 & -0.360 \\ 0 & 0 \\ 0 & 2.230 \end{bmatrix}.$$

Deriving a closed-form minimum-time control law for a system of such dimensions is a daunting and inextricable task. The method of this paper provides a generic and computational alternative to simply obtain the control sequence. The control constraint set is a ball of the form $\mathcal{U} = \{u \in \mathbb{R}^2 : \|u\|_2 \leq 1\}$. Starting the system in the state $x_0 = [10 \ -10 \ 5]^\top$ and applying our method with a control horizon $T = 10$ and a linearly increasing weight sequence, we obtain the control and state trajectories plotted in Figure 3. Note that the control tends to take extreme values (i.e., such that $\|u\|_2 = 1$) when the state is far from zero. The values of the control become smaller when the state approaches zero. This observation is illustrated in Figure 4.

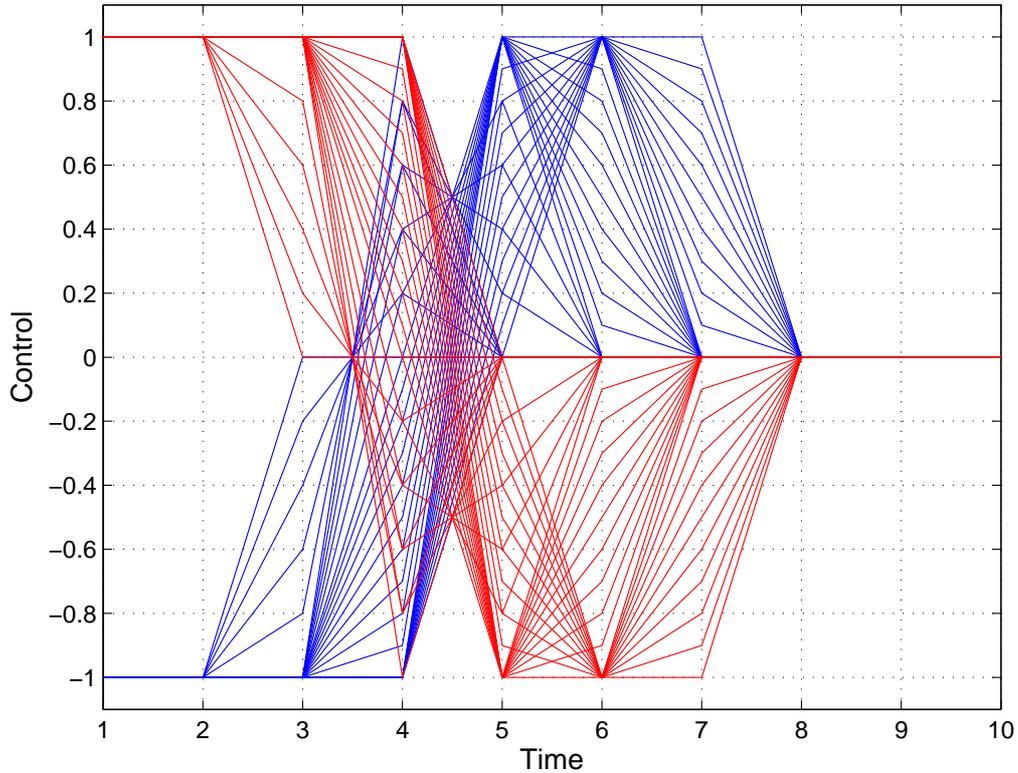

Figure 2: Optimal control trajectories over time samples.

## 5 Conclusion

We have presented a numerical but quite practical solution to the minimum-time control problem for discrete-time LTI systems. The optimal control sequence is efficiently obtained via a sparse optimization strategy. We showed that, under certain conditions, the time-optimal control can be exactly computed using convex optimization techniques for which reliable numerical solvers exist. The main advantage of our approach is that it is very generic in the sense that it can be straightforwardly applied to an arbitrary linear system with more than one input. An interesting open issue would be to consider the situations where the state has to be driven to zero in minimum-time under the constraint that the state trajectory must pass through some pre-specified points. Another research path for future work is to extend the proposed method to minimum-time problems involving somewhat arbitrary target sets.

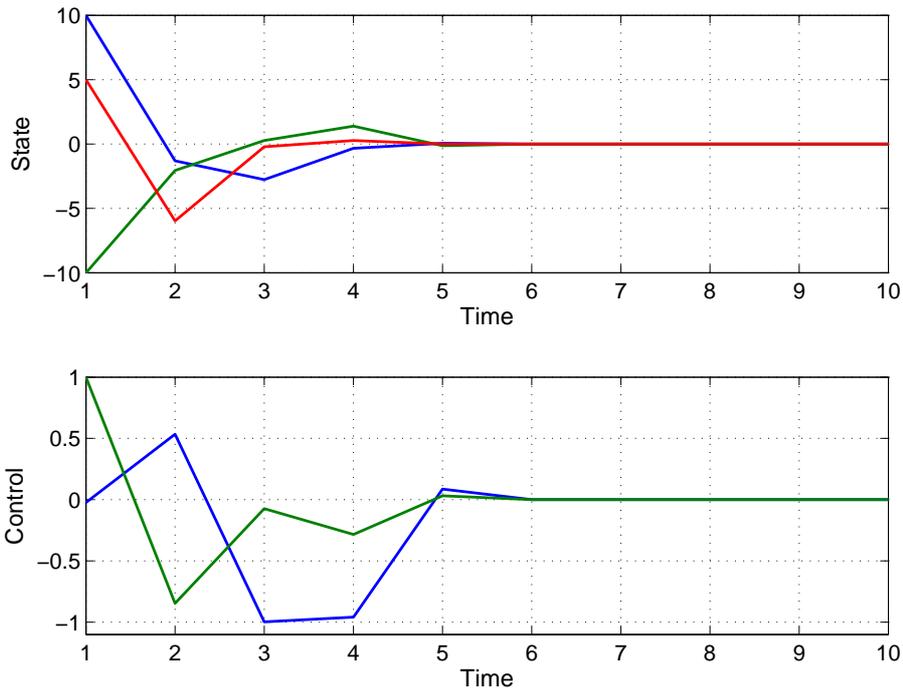

Figure 3: State and control trajectories over time samples when $x_0 = [10 \ -10 \ 5]^\top$.

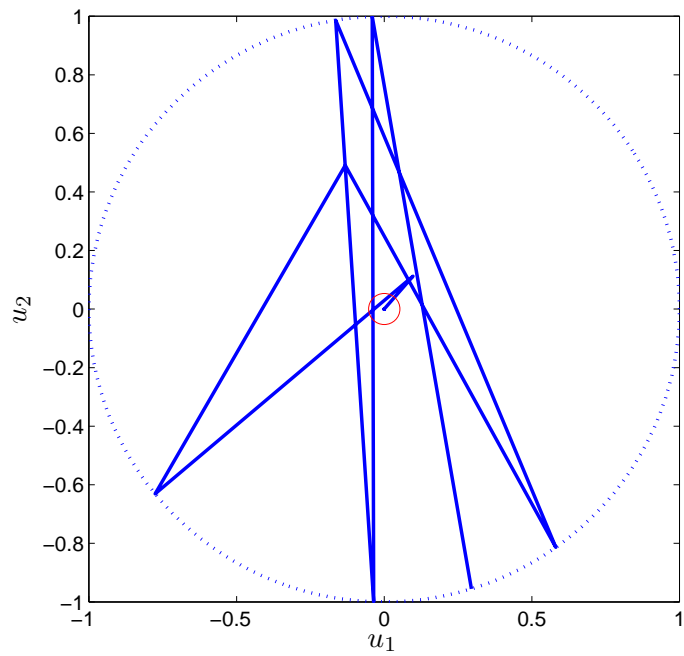

Figure 4: Optimal control trajectory for $x_0 = [50 \ -50 \ -50]^\top$.